\documentclass{iaus}
\usepackage{graphicx}

\title[IAUS 265.~~Fe I/Fe II ionization equilibrium: NLTE versus LTE]
{Fe I/Fe II ionization equilibrium \\in cool stars: NLTE versus LTE}

\author[Mashonkina et al.]
{Lyudmila Mashonkina$^{1,2}$, Thomas Gehren$^2$, Jianrong Shi$^3$, \\ Andreas Korn$^4$, \and Frank Grupp$^2$}

\affiliation{$^1$ Institute of Astronomy, Russian Academy of Science, \\ Pyatnitskaya 48, 119017 Moscow, Russia \\ email: {\tt lima@inasan.ru} \\[\affilskip]
$^2$ Universit\"ats-Sternwarte M\"unchen, \\ Scheinerstr. 1, 81679 M\"unchen, Germany \\ email: {\tt lyuda, gehren, fug@usm.lmu.de}\\[\affilskip]
$^3$ National Astronomical Observatories, Chinese Academy of Sciences, \\A20 Datun Road, Chaoyang District, Beijing 100012, PR China  \\ email: {\tt sjr@bao.ac.cn}\\[\affilskip]
$^4$ Department of  Physics and Astronomy,
Uppsala University, \\ Box 515, 75120 Uppsala, Sweden\\ email: {\tt andreas.korn@fysast.uu.se}}

\pubyear{2009}
\volume{265}  
\pagerange{1--4}
\jname{Chemical Abundances in the Universe: Connecting First Stars to Planets}
\editors{K. Cunha, M. Spite \& B. Barbuy, eds.}

\newcommand{\Teff}{T_{\rm eff}}
\newcommand{\logg}{\rm log~ g}
\newcommand{\Vmic}{V_{\rm mic}}
\newcommand{\kms}{km\,s$^{-1}$}
\newcommand{\eps}[1]{\log\varepsilon_{\rm #1}}

\begin{document}

\maketitle

\begin{abstract}
Non-local thermodynamic equilibrium (NLTE) line formation for neutral and
singly-ionized iron is considered through a range of stellar parameters
characteristic of cool stars. A comprehensive model atom for Fe~I and Fe~II is
presented. Our NLTE calculations support the earlier conclusions that the
statistical equilibrium (SE) of Fe~I shows an underpopulation of Fe~I terms.
However, the inclusion of the predicted high-excitation levels of Fe~I in our
model atom leads to a substantial decrease in the departures from LTE.
As a test and first application of the  Fe~I/II model atom, iron abundances are
determined for the Sun and four selected stars with well determined stellar
parameters and high-quality observed spectra. Within the error bars,
lines of Fe~I and Fe~II give consistent abundances for the Sun and two
metal-poor stars when inelastic collisions with hydrogen atoms are taken into
account in the SE calculations. For the close-to-solar metallicity stars Procyon
and $\beta$~Vir, the difference (Fe~II - Fe~I) is about 0.1~dex independent of
the line formation model, either NLTE or LTE. We evaluate the influence of
departures from LTE on Fe abundance and surface gravity determination for cool
stars.

\keywords{atomic data, line: formation, stars: atmospheres}
\end{abstract}

\firstsection 
\section{Introduction}

Iron plays an outstanding role in studies of cool stars thanks to quite
numerous lines in the visible spectrum, which are easy to detect even in very
metal-poor stars. Iron serves as a reference element for all astronomical
research related to stellar nucleosynthesis and chemical evolution of the
Galaxy. Iron lines are used to determine
the surface gravity, $\logg$, and the microturbulence $\xi$ of
stellar atmospheres. In the atmosphere with $\Teff > 4500$~K, neutral
iron is a minority species. The ionization equilibrium between Fe~I
and Fe~II and the excitation equilibrium of Fe~I easily deviate from
thermodynamic equilibrium. Since the beginning of the 1970s a
number of studies attacked the problem of non-local thermodynamic
equilibrium (NLTE) for Fe (e.g., \cite[Athay \& Lites (1972)]{Athay1972},
\cite[Thevenin \& Idiart
(1999)]{Thevenin1999}, \cite[Gehren \etal\ (2001)]{Gehren2001}). However, a
consensus on the expected magnitude of the NLTE effects was not reached.

In this study, we update the model atom of Fe I-II treated by \cite[Gehren
\etal\ (2001)]{Gehren2001} (hereafter Paper~I) and apply it to analysis of the
Fe spectrum in the Sun and selected cool stars with the aim of
empirically constraining the role of inelastic collisions with hydrogen
atoms in the SE of Fe~I-II.

\firstsection

\section{The Fe model atom}

In all previous NLTE calculations, the model atom of Fe~I was build
using measured energy levels. The experimental analysis of \cite[Nave
\etal\ (1994)]{Nave1994} with later updates provided 965 energy levels for Fe~I.
A comparison with the calculated Fe~I atomic structure (\cite[Kurucz
(2007)]{Kurucz2007}) reveals that the system of measured levels is nearly
complete below excitation energy, $E_{\rm exc}$, 5.6~eV,
however, laboratory experiments do not see most of the high-excitation
levels with $E_{\rm exc} >$
7.1~eV. As already shown in the first NLTE studies, the main NLTE
mechanism for Fe~I is the overionization of low-excitation levels by
ultraviolet radiation. The role of high-excitation levels is
to compensate, in part, for population losses via collisional coupling
to the large continuum reservoir, with subsequent spontaneous
transitions down to low-excitation levels. Therefore, the system of
levels in the model atom of Fe~I has to be fairly complete at least up to 0.5~eV
(mean kinetic energy of electrons in the atmosphere) below the ionization limit.

For Fe~I, our model atom was constructed using all known energy levels and the
predicted levels with $E_{\rm exc}$ up to 7.83~eV, in total, 2970 levels.
Multiplet fine structure was neglected for all terms. The predicted and measured
levels with close energies were combined resulting in 233
terms. In addition, six super-levels were made up from the remaining
predicted levels. For 11958 radiative transitions occurring in this
atom of Fe~I, $gf$-values were taken from the \cite[Nave \etal\
(1994)]{Nave1994} compilation, where available, and \cite[Kurucz
(2007)]{Kurucz2007} calculations. Photoionization cross-sections of the IRON
project (\cite[Bautista (1997)]{Bautista1997}) have been used for 149 levels and
a hydrogenic approximation for the remaining levels. The collisional rates were
computed as in Paper~I.

For Fe~II, we rely on the reference model atom treated in Paper~I. In
this study, it was reduced and includes now the levels with $E_{\rm exc}$ up to
10~eV. The main uncertainty of the NLTE calculations for Fe~I and II is
the treatment of the poorly known inelastic collisions with hydrogen atoms. We
employ the formula of \cite[Steenbock \&  Holweger (1984)]{Steenbock1984} for
allowed transitions and a simple correlation between hydrogen and electron
collisional rates,  $C_H = C_e \sqrt{(m_e/m_H)} N_H/N_e$, for forbidden
transitions. Calculations were performed with the hydrogen collision enhancement
factor $S_{\rm H}$, which was varied between 0 and 3.

\firstsection

\section{Results}

The coupled radiative transfer and statistical equilibrium equations are solved
with an improved version of the DETAIL program (\cite[Butler \&
Giddings (1985)]{detail}) based on the accelerated lambda iteration. All
calculations are performed with plane-parallel, homogeneous, LTE, and blanketed
model atmospheres computed with the MAFAGS-OS code (\cite[Grupp \etal\
(2009)]{Grupp09}).

For comparison with observed data, a total of 43 lines of Fe~I and 18 lines of
Fe~II were chosen. For the Sun and HD\,84937, the analysis was extended
to a larger line list including 271 lines of Fe~I and 34 lines of Fe~II. The
Sun is also used as a reference star for a line-by-line differential
analysis of stellar spectra. Solar flux observations were taken from the Kitt
Peak Solar Atlas (\cite[Kurucz \etal\ (1984)]{Atlas}). The absolute
solar iron abundances were determined using $gf$-values from \cite[O'Brian et
al. (1991)]{fe-OWL91} and \cite[Mel\'endez \& Barbuy (2009)]{fe-MB09} for Fe~I
and Fe~II, respectively. We find that virtually all models of line formation,
whether LTE or NLTE with $S_H \ge 0.1$, lead to acceptable solar ionization
equilibria within their $1\sigma$ error bars. To show the maximal NLTE effect on
abundance determination, Table\,\ref{startab} presents the average abundances
for both ionization stages derived from the NLTE with $S_H$ = 0 (denoted as
NLTE$_0$) and LTE calculations.

\begin{table}
  \begin{center}
\caption{Stellar parameters and iron abundances obtained
 for selected stars}
\label{startab}
\begin{tabular}{rclccccc}
\noalign{\smallskip} \hline \noalign{\smallskip}
 HD & $\Teff$ & $\logg$ & $\Vmic$, & \multicolumn{2}{c}{ [Fe/H]$_{\rm I}$} & \multicolumn{2}{c}{[Fe/H]$_{\rm II}$} \\
    &         &          &  \kms &  NLTE$_0$ & LTE      &  NLTE$_0$ & LTE             \\
\noalign{\smallskip} \hline \noalign{\smallskip}
   Sun & 5777 & 4.44                       & 0.9 & ~~7.63\scriptsize{$\pm$0.08} & ~~7.49\scriptsize{$\pm$0.10}~~~~ & ~~7.44\scriptsize{$\pm$0.06} & ~~7.45\scriptsize{$\pm$0.06} \\ %
 10700 & 5377 & 4.53                       & 0.8 & $-$0.43\scriptsize{$\pm$0.04} & $-$0.49\scriptsize{$\pm$0.02}~~~~ & $-$0.53\scriptsize{$\pm$0.05} & $-$0.52\scriptsize{$\pm$0.05} \\
 61421 & 6510 & 3.96\scriptsize{$\pm$0.02} & 1.8 & $-$0.10\scriptsize{$\pm$0.06} & $-$0.14\scriptsize{$\pm$0.05}~~~~ & $-$0.04\scriptsize{$\pm$0.03} & $-$0.04\scriptsize{$\pm$0.03} \\
 84937 & 6350 & 4.00\scriptsize{$\pm$0.12} & 1.7 & $-$1.94\scriptsize{$\pm$0.06} & $-$2.16\scriptsize{$\pm$0.07}~~~~ & $-$2.08\scriptsize{$\pm$0.04} & $-$2.11\scriptsize{$\pm$0.04} \\
102870 & 6060 & 4.11\scriptsize{$\pm$0.01} & 1.4 & ~~0.04\scriptsize{$\pm$0.03} & ~~0.04\scriptsize{$\pm$0.03}~~~~ & ~~0.13\scriptsize{$\pm$0.04} & ~~0.12\scriptsize{$\pm$0.04} \\
\noalign{\smallskip} \hline \noalign{\smallskip}
\end{tabular}
 \end{center}
\end{table}

\begin{figure}[b]
\begin{center}
 \includegraphics[width=4in]{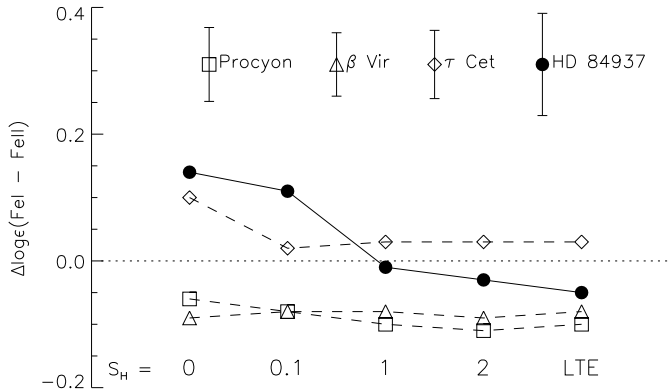}
 \vspace*{-1.0 cm}
 \caption{The difference in abundance between Fe~I and Fe~II in
  selected stars from calculations with various line formation models.
  For each star, the error bars is indicated in the upper part of panel.}
   \label{fig1}
\end{center}
\end{figure}

Four stars with effective temperature and surface gravity measured from the
model-independent methods were chosen to investigate the ionization equilibrium
between Fe~I and Fe~II for various $S_H$ values. They are listed in
Table\,\ref{startab} together with the $\Teff$ and $\logg$ values taken from
\cite[Di Folco et al. (2004)]{tauCet} for HD\,10700 ($\tau$~Cet), \cite[Allende
Prieto \etal\ (2002)]{procyon} for HD\,61421 (Procyon), \cite[Korn \etal\
(2003)]{Korn03}  for HD\,84937, \cite[North \etal\ (2009)]{betaVir}  for
HD\,102870 ($\beta$~Vir). Observational data were obtained with the
FOCES spectrograph at the 2.2m telescope of the Calar Alto Observatory
during a number of observing runs between 1997 and 2005, with a spectral
resolution of $R \simeq$ 60\,000 and a signal-to-noise ratio $S/N \ge 200$.

The NLTE, $S_H$ = 0 and LTE abundances obtained from the lines of Fe~I (denoted
as [Fe/H]$_{\rm I}$) and Fe~II ([Fe/H]$_{\rm II}$) are presented in
Table\,\ref{startab}. It is worth noting that, with the updated model atom of
Fe~I-II, the departures from LTE are substantially smaller compared to those
from the previous studies. For example, with $S_H$ = 0, we obtain an
average NLTE abundance correction $\Delta_{\rm NLTE} = \eps{NLTE}-\eps{LTE}$ =
0.22~dex for the Fe~I lines in HD\,84937, while the corresponding value
amounts 0.40~dex in \cite[Korn \etal\ (2003)]{Korn03}.
Figure\,\ref{fig1} displays the abundance difference between Fe~I and Fe~II for
various assumptions for the hydrogen collisions. We find that
NLTE with pure electronic collisions ($S_H$ = 0) is not acceptable for HD\,84937
and $\tau$~Cet. This indicates the need for thermalizing processes not involving
electrons in the atmosphere of metal-poor stars. For each object, the NLTE
effect on abundance determination is small (within the error bars) when hydrogen
collisions are included with $S_H \ge 1$. For Procyon and $\beta$~Vir, the mean
Fe abundance from Fe~I lines is about 0.1~dex lower compared to that from  Fe~II
lines. The origin of such a discrepancy will be investigated in a
forthcoming paper. For the Fe~I/Fe~II ionization equilibrium in two metal-poor
stars, LTE seems to be as good as NLTE with $S_H \ge 1$.

The NLTE calculations were performed with $S_H = 1$ for the small grid of model
atmospheres with $\Teff =$ 5000 and 6000~K, [M/H] = $-1$ and $-3$, and $\logg$
ranging between 2 and 4 in order to inspect the departures from LTE depending on
stellar parameters. Negligible NLTE effects were obtained for Fe~II. Fe~I is
subject to significant NLTE effects for low gravity ($\logg < 3$) and very
metal-poor models. An important consequence is that surface gravities of giants
and very metal-poor stars derived by LTE analysis are in error with a magnitude
strongly depending on $\logg$/[Fe/H]. For example, LTE leads to a 0.26~dex lower
gravity for $\Teff =$ 5000~K, $\logg$ = 2, and [Fe/H] = $-3$.

\firstsection

\section{Conclusions}

\begin{itemize}
\item Completeness of model atom for Fe~I is important for a correct
calculation of the Fe~I/Fe~II ionization equilibrium in the atmosphere of cool
stars. \item Thermalizing processes not involving electron collisions have to be
included in the SE calculations for Fe~I-II. Collisions with hydrogen atoms
could be good candidates for such processes. \item Fe I is affected by
significant NLTE effects for giants and very metal-poor stars. \item Only minor
departures from LTE are obtained for Fe II.
\end{itemize}

\bigskip

{\it Acknowledgements.} L.M. acknowledges a partial support from the
International Astronomical Union, the Russian Foundation for Basic Research
(08-02-92203-GFEN), and the Russian Federal Agency on Science and
Innovation (02.740.11.0247) of the participation at the IAU XXVII General
Assembly. This study is supported by the Deutsche Forschungsgemeinschaft (GE 490/34.1). A.K. acknowledges support by the Swedish Research Council (VR).

\end{document}